\documentclass[epj,final]{svjour}
\usepackage{graphics}
\usepackage{psfig}
\usepackage{amssymb}
\begin{document}
\title{Vertically shaken column of spheres. Onset of fluidization}

\author{Simon Renard\inst{1} \and Thomas Schwager\inst{1} \and
  Thorsten P\"oschel\inst{1}\and Clara Salue\~na\inst{2} }

\institute{
  \inst{1} Humboldt-Universit\"at, Charit\'e, Institut f\"ur Biochemie, 
  Monbijoustra{\ss}e 2, D-10117 Berlin, Germany\\ 
  http://summa.physik.hu-berlin.de/$\sim$kies/\\
  \inst{2} Humboldt-Universit\"at, Institut f\"ur Physik,
  Invalidenstra{\ss}e 110, D-10115 Berlin, Germany} \date{Received:
  \today / Revised version: }

\abstract{The onset of surface fluidization of granular material in a
  vertically vibrated container, $z=A\cos\left(\omega t\right)$, is
  studied experimentally. Recently, for a column of spheres it has
  been theoretically found~\cite{BelowG} that the particles lose
  contact if a certain condition for the acceleration amplitude
  $\ddot{z} \equiv A\omega^2/g = f(\omega)$ holds. This result is in
  disagreement with other findings where the criterion
  $\ddot{z}=\ddot{z}_{\rm crit}=\mbox{const.}$ was found to be the
  criterion of fluidization. We show that for a column of spheres a
  critical acceleration is not a proper criterion for
  fluidization and compare the results with theory.\\
}
\maketitle

\section{Introduction}
Granular material confined in a vertically vibrating container reveals
complex effects such as surface structure formation~(e.g.
\cite{Coulomb:1773,MeloUmbanhowarSwinney:1994}), spontaneous heap
formation (e.g. \cite{DinkelackerHueblerLuescher:1987}), convection
(e.g. \cite{EhrichsJaegerKarczmarKnightKupermanNagel:1994}) and
others. A precondition common to all these effects is that (at least)
the particles at the free surface of the granular material lose
contact to their neighbours for at least a small part of the
oscillation period. If a granular material is agitated in a way so
that the particles at the surface separate from their neighbours we
will call the material fluidized.
 
The conditions under which surface fluidization occurs are not clear
yet. For sinusoidal vertical excitation of the container most of the
literature reports that surface fluidization starts as soon as the
acceleration amplitude $A\omega^2$ of the oscillation $z=A\cos\omega
t$ exceeds gravity $g$ or another critical constant. The Froude number
is defined as $\Gamma=A\omega^2/g$ and, hence, it has been reported in
numerous publications that surface fluidization or, respectively,
effects which require fluidization, occur for $\Gamma>1$.

Few references \cite{BarkerMehta:1993a} report, however, that in
numerical simulations surface fluidization was observed for
$\Gamma\lesssim 1$. So far these results have not been confirmed
experimentally.

Whereas a rigid solid body, e.g. a single sphere on a oscillating
surface, would certainly start to jump if the acceleration amplitude
$A\omega^2$ exceeds gravity $g$, there are arguments which may lead to
a different conclusion for a shaken amount of granular material, i.e.,
a many body system:
\begin{itemize}
\item The deformation-force-law of contacting bodies may be nonlinear
  due to geometrical effects, even if the material deformation is
  small enough to assume linear material properties, i.e., Hooke's
  law. For contacting ideally elastic spheres it has been derived by
  Hertz \cite{Hertz:1882} that the interaction force $F$ scales with
  the deformation $\xi$ as $F\sim \xi^{3/2}$. This law applies for all
  particle contacts (under mild conditions) provided the deformation
  $\xi$ is small enough \cite{Brilliantov}.  As a result, nonlinear
  phenomena like the propagation of solitary waves has been shown to
  appear in bead chains with Hertz contact law
  \cite{CosteFalconFauve:1994PRE}.
  
\item Under the influence of gravity the particles are deformed
  differently even at rest, depending on their vertical position in
  the material. Therefore, the speed of sound in the material is not
  uniform but a function of the vertical coordinate. This property may
  lead to complicated pulse motion through the material and has been
  reported also for chains of contacting spheres
  \cite{SinkovitsSen:1995}.
 
\item In a polydisperse granular material the geometrical properties,
  i.e., the contact network, varies with the applied force. If the
  material is loaded the particles are deformed and more and more
  contacts emerge. This effect leads to a more complicated deformation
  -- force law, $F\sim\xi^\gamma$ with $1.5 \le \gamma \le 4$
  \cite{HerrmannStaufferRoux:1987a}.
\end{itemize}

In a recent theoretical paper \cite{BelowG} the granular material was
modeled as a vertically shaken column of viscoelastic spheres. The
main result of \cite{BelowG} is that the column may be fluidized even
if the condition $\Gamma>1$ does not hold. Instead a different
condition for fluidization was derived: Assuming viscoelastic material
properties the force between two contacting spheres of radius $R$ at
vertical positions $z_k$ and $z_{k+1}$ reads \cite{BSHP}
\begin{equation}
  \label{eq:BSHP}
  F_{k,k+1} = -\sqrt{R} \left(\mu\xi_{k,k+1}^{3/2}+\alpha\dot{\xi}_{k,k+1}
\sqrt{\xi_{k,k+1}}\right)\,,
\end{equation}
where $\xi_{k,k+1}\equiv 2R-\left|z_k-z_{k+1}\right|$ is the
compression and $\mu$ and $\alpha$ are elastic and dissipative
material constants (details see \cite{BSHP}). The height of the column
of $N$ spheres is $L=2NR$. It was found that the top sphere separates
from its neighbour if
\begin{equation}
  \label{eq:flui}
  \frac{A\omega^2}{g}>\Gamma\left(\frac35\right)\left|M^{2/5}
\,J_{-2/5}\left(2M\right)\right|\,,
\end{equation}
where $J_{-2/5}$ is the (complex) Bessel function and $M$ abbreviates
\begin{equation}
  M\equiv\frac{(18\pi)^{1/3}}{5}\left(\frac{\mu L^5\rho^2}
{g}\right)^{1/6}\frac{\omega}{\sqrt{3\mu-2i\omega\alpha}}\,,
\end{equation}
with $i=\sqrt{-1}$ and $g$ and $\rho$ being gravity and material
density.

For small frequency $\omega$ the Taylor expansion of the rhs. of the
condition (\ref{eq:flui}) yields
\begin{equation}
  \frac{A\omega^2}{g}>1-B_2\omega^2+B_4\omega^4
  \label{eq.flui.small}
\end{equation}
with
\begin{eqnarray*}
  B_2&\equiv&\frac{18^\frac{2}{3}}{45}
  \left(\frac{\pi^2\rho^2L^5}{g\mu^2}\right)^\frac{1}{3}\\
  B_4&\equiv&\frac{1}{18^\frac{2}{3}}\! 
  \left(\frac{\pi^2\rho^2L^5}{g\mu^2}\right)^\frac{2}{3}\!\!\!
\left(\!\frac{3}{100}+\frac{4\cdot 324^\frac{2}{3}}{405}
  \frac{\alpha^2}{\pi^2} \! \left(\frac{g \pi^4}{\mu^4 \rho^2 L^5}
\right)^\frac{1}{3}\!\right)
\end{eqnarray*}
Both coefficients $B_2$ and $B_4$ are positive definite values, i.e.,
for {\em all} materials there is a range of frequency where the rhs.
of the inequality (\ref{eq:flui}) is smaller than one. This means that
there exists always a frequency interval where the column fluidizes
although the acceleration amplitude of shaking is smaller than
gravity, $A\omega^2/g <1$. The full derivation of the sketched theory
and the discussion of the frequency range where the condition
(\ref{eq:flui}) holds can be found in Ref. \cite{BelowG}. For our
experimental system consisting of steel spheres, of density
$\rho$=7700 Kg/m$^3$, Young modulus $Y$=210 GPa, Poisson ratio
$\nu=0.29$, in a column of $L=$0.6 m, the Taylor expansion Eq.
(\ref{eq.flui.small}) holds for frequencies $\omega$ much smaller than
1000 s$^{-1}$, which covers entirely the experimentally accessible
range of frequencies.  \medskip

In the present paper we want to report experimental results on the
onset of fluidization of a vertically shaken column of spheres, i.e.,
on the same system which was studied theoretically in Ref.
\cite{BelowG}.  The dynamical behaviour of a vibrated one dimensional
system of spheres in the well fluidized regime was studied e.g.
in~\cite{BernuDelyonMazighi:1994,x1}. This regime is explicitely not
considered here. In~\cite{x} a column of particles which interact as
linear springs with linear damping,
$\ddot{\xi}+2\mu\dot{\xi}+\Omega^2\xi=0$, was studied with Molecular
Dynamics. For their sets of parameters the authors found that
fluidization occurs above $\Gamma\approx 1.1$, however, the authors
claim that the onset of fluidization depends on the choice of the
parameters $\Omega$ and $\mu$. Within their model it might be possible
to get also fluidization for $\Gamma<1$, depending on the material
parameters. If one considers spheres, however, in the limit of
vanishing damping the Hertz law implies that the duration of contact
behaves as $t_c\sim g^{-1/5}$, whereas for the model considered in
\cite{x} one finds a constant $t_c=\mbox{const.}=\pi/\Omega$. Hence,
the linear spring model certainly does not describe the contact of
spheres, not even in the case of pure elastic contact. For the same
linear spring system in \cite{xx} the detachment effect was reported,
the same effect for viscoelastic spheres was discussed in \cite{xxx}.

\section{Experimental setup}\label{sec:setup}

We expect that the effect to be measured is quite small, i.e., for the
critical parameters of shaking when the material starts to fluidize we
expect $1-A\omega^2/g$ to be a small value. Therefore, we have to
adjust the amplitude of shaking $A$ and in particular the frequency
$\omega$ with good accuracy. Moreover, we have to assure that no
energy originating from different frequencies than the driving
frequency enters the column of spheres, e.g., due to high frequency
noise.

\begin{figure}[htbp]
\begin{minipage}{8.5cm}
\centerline{\psfig{figure=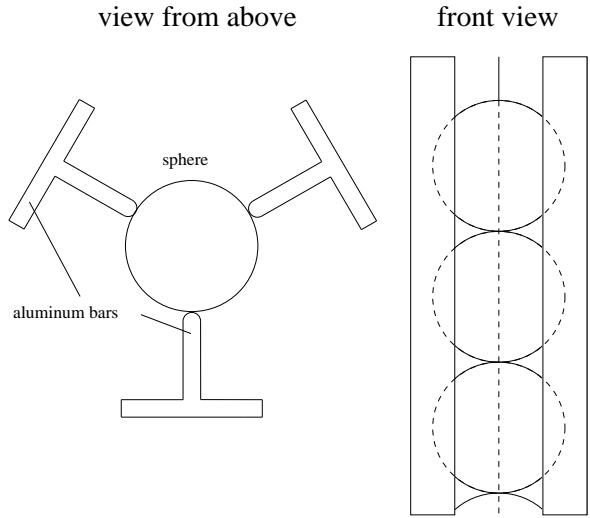,width=8cm,angle=0}}
\psfull
    \caption{The column of spheres moves in a framework of aluminum bars}
    \label{fig:frame}
\end{minipage}
\end{figure}

The column of spheres was confined in a framework of three aluminum
bars, which assures pure one dimensional motion of the spheres at low
friction, see Fig.~\ref{fig:frame}.

\begin{figure}[htbp]
\begin{minipage}{8.5cm}
\centerline{\psfig{figure=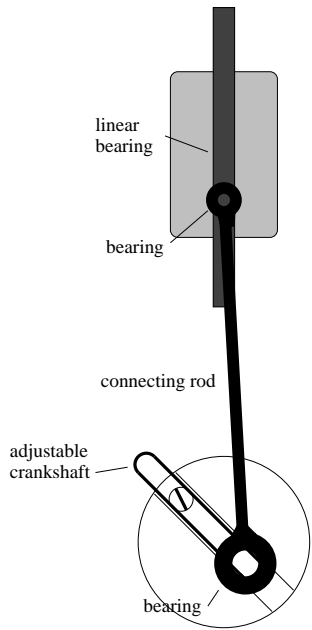,width=4cm,angle=0}}
    \caption{The framework which holds the column of spheres was 
      mounted on a linear bearing and was driven via a crankshaft.}
    \label{fig:pleuel}
\end{minipage}
\end{figure}

The framework with the spheres was mounted on a precisely vertically
balanced linear bearing which was driven by a stepping motor via a
crankshaft with adjustable eccentricity, Fig.~\ref{fig:pleuel}.

\begin{figure}[htbp]
\begin{minipage}{8.5cm}
\centerline{\psfig{figure=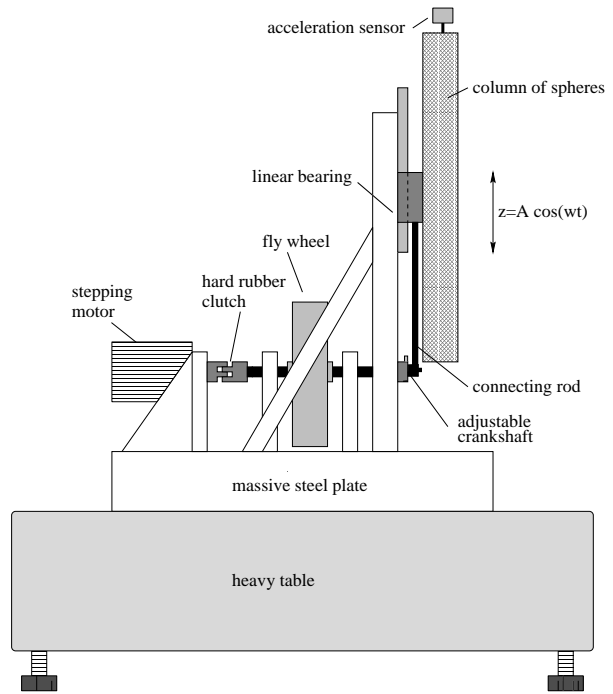,width=8cm,angle=0}}
    \caption{Sketch of the experimental device}
    \label{fig:gesamtanlage_vorn}
\end{minipage}
\end{figure}

\begin{figure}[htbp]
\begin{minipage}{8.5cm}
\centerline{\psfig{figure=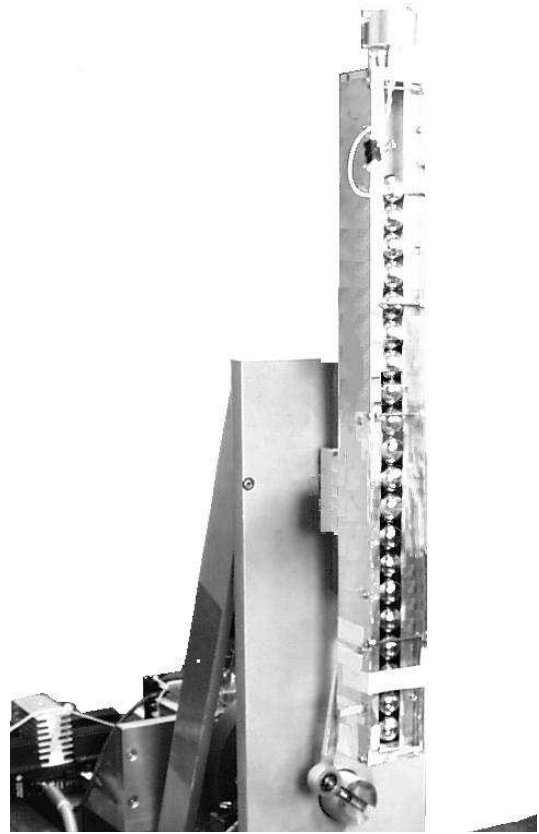,width=7cm,bblly=93pt,bbllx=174pt,bbury=425pt,bburx=383pt,angle=0,clip=}}
%\centerline{\psfig{figure=figs/foto_repaint.eps,width=7cm,bblly=68pt,bbllx=22pt,bbury=817pt,bburx=503pt,angle=0,clip=}}
\caption{Experimental device with column of steel spheres}
\label{foto9}
\end{minipage}
\end{figure}

The motor was computer controlled, i.e., in precisely 20,000 impulses
the motor axis revolves once. This high angular resolution provides
quasi-steady motion. The motion was smoothened additionally using a
flywheel (with diameter 160~mm, thickness 32~mm, mass 4.5~kg and
moment of inertia $J=0.0144 ~\mbox{kg\,m}^2$) which was fixed on the
axis and a hard rubber clutch connecting the motor axis and the axis
of the shaking device (see Fig.~\ref{fig:gesamtanlage_vorn}). Using an
acceleration sensor we checked that the finite step size per computer
signal does not influence the sinusoidal motion of the column of
spheres. The entire mechanical device was mounted on a massive
oscillation damping table. Figure~\ref{fig:gesamtanlage_vorn} sketches
the device and Fig.~\ref{foto9} shows a photograph.

In the experiments the column consists of up to 20 ball bearing balls
of radius $R=12.5$ mm. There are several possibilities to determine
the onset of fluidization, i.e., for fixed amplitude $A$ the critical
frequency $\omega$ when the top sphere separates from its neighbour:
\begin{itemize}
\item Direct observation: a comoving camera was mounted to the probe
  carrier and a LASER pointer was fixed at the opposite side of the
  column. As soon as fluidization starts at the critical frequency the
  emerging submillimeter gap between the top particle and its
  neighbour can be recognized as a flash on the screen.
\begin{figure}[htbp]
\begin{minipage}{8.5cm}
\centerline{\psfig{figure=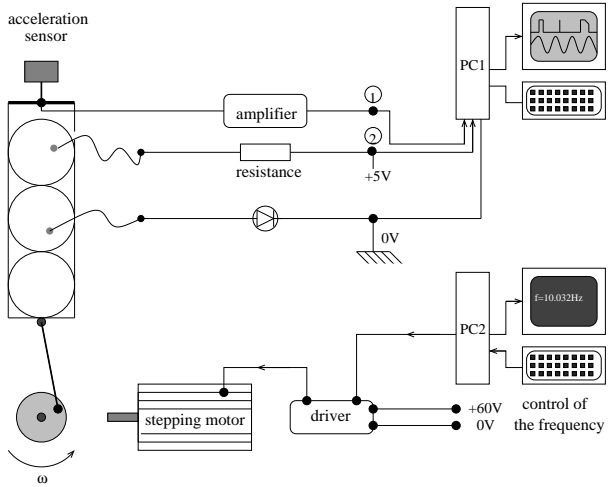,width=8cm,angle=0}}
\psfull
    \caption{The onset of fluidization was determined by observing 
      interruptions of the electrical circuit.}
    \label{fig:LASER}
\end{minipage}
\end{figure}

\item Acoustic method: When fluidization occurs there emer\-ges a
  periodic acoustic signal originating from the collision of the top
  particles. This signal can be detected either using a microphone or
  the acceleration sensor which was mounted on the top of the device,
  see Fig.~\ref{fig:gesamtanlage_vorn}.
  
\item Electric circuit: We attached a very thin wire (diameter
  $<0.1$\,mm) to the top particle and its neighbour and observed the
  contact of the spheres using an oscilloscope.

\end{itemize}

We checked all three methods and found that the results are very
similar. We have chosen the electric method since it turned out to be
the most simple and most reliable one. The direct observation requires
complicated mechanical tuning and for the acoustic method one has to
filter the periodic signal from other sources of sound as, e.g., the
sound of the motor.

\section{Raw Data}

\begin{figure}[htbp]
  \centerline{\psfig{figure=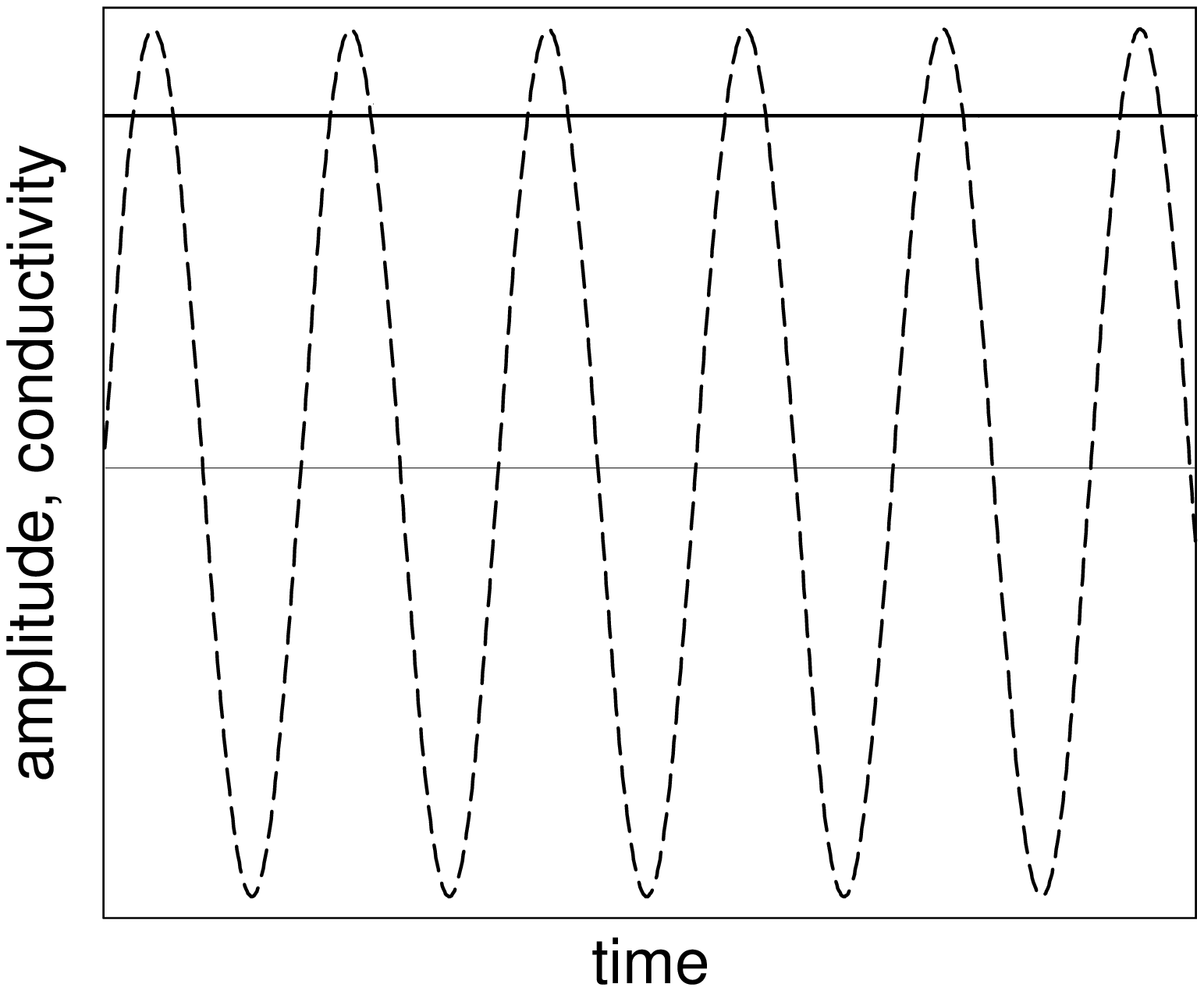,width=5cm}}
  \centerline{\psfig{figure=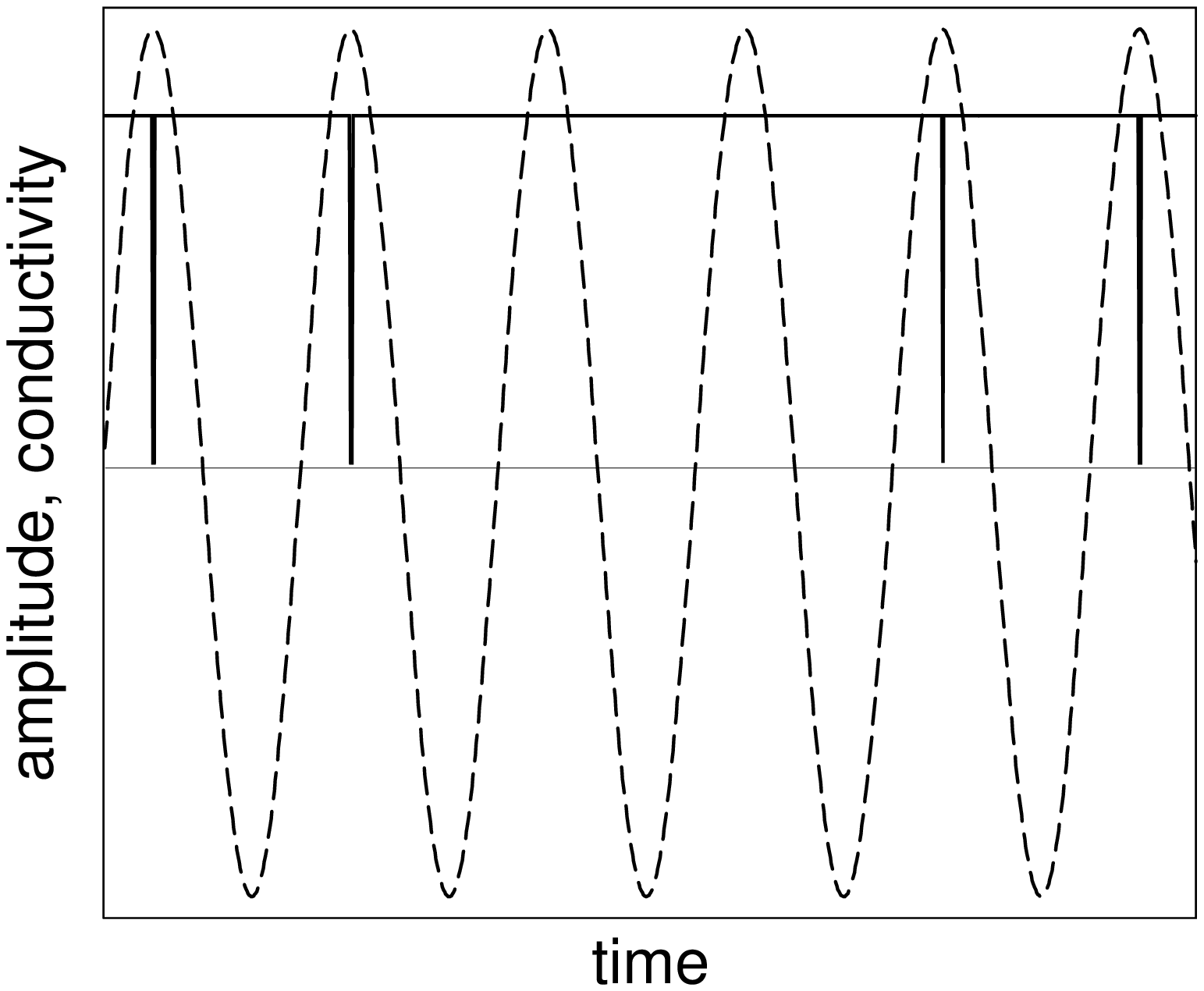,width=5cm}}
  \centerline{\psfig{figure=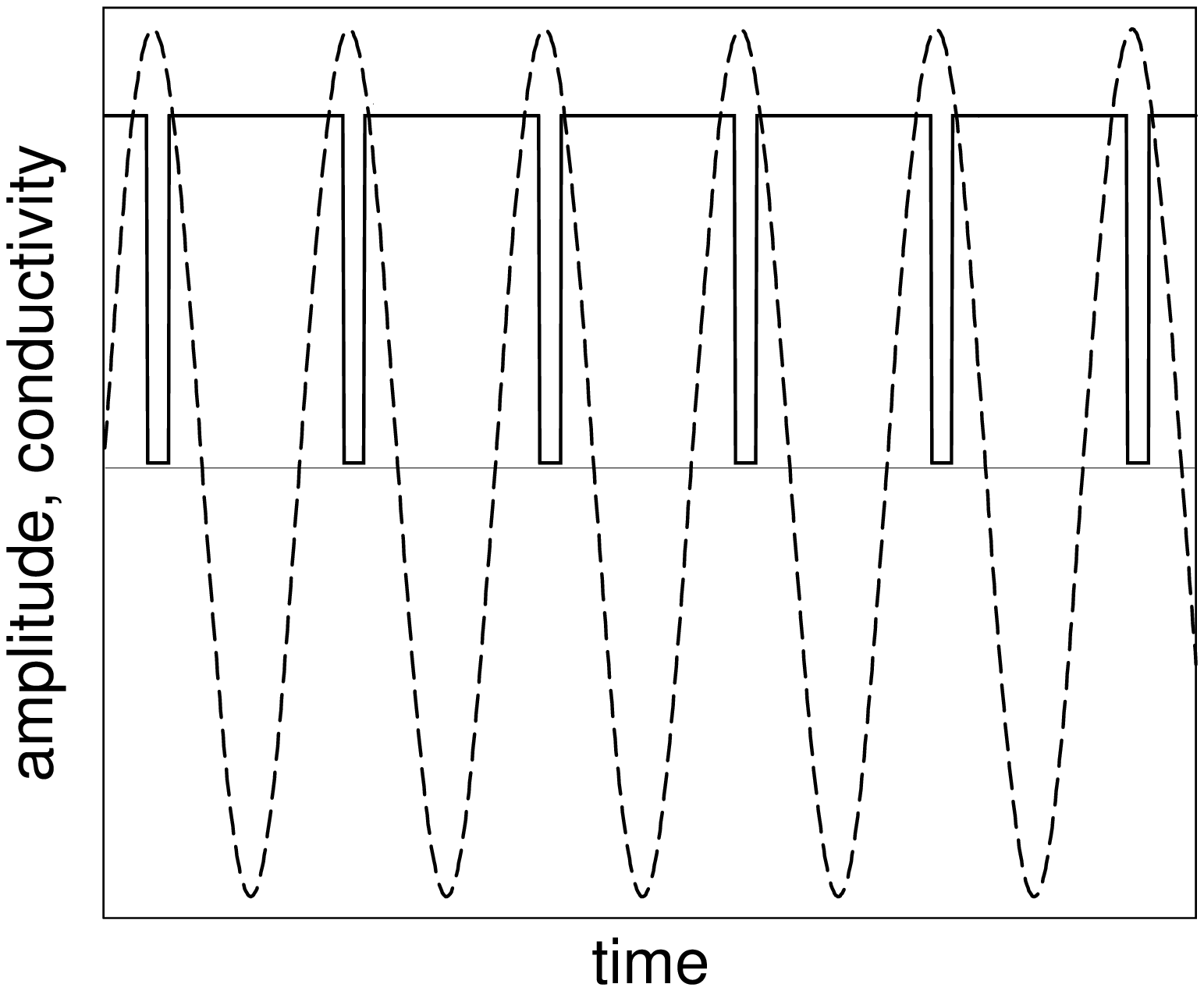,width=5cm}}
  \caption{For subcritical shaking the particles are always in contact, 
    i.e., the conductivity (solid line) is constant (top). For
    over-critical shaking the conductivity drops to zero in each
    period (bottom). For critical values the conductivity shows
    irregular peaks (middle) }
  \label{fig:cond}
\end{figure}

For fixed amplitude $A$ we want to determine the according smallest
frequency $\omega$ for which the top sphere separates from its
neighbour in each period. Figure \ref{fig:cond} shows a sketch of the
oscilloscope signal for subcritical (top), critical (middle) and
over-critical frequency, all for the same amplitude of shaking. For
subcritical frequency the top particle follows the motion of the
vibrating table resulting in a sinusoidal signal of the acceleration
sensor and a horizontal line for the conductivity (limited by a
resistor). At over-critical frequency the conductivity drops to zero
for a certain time interval in each period. Close to the critical
frequency the electrical contact between the spheres is not in each
period interrupted, instead we get a noisy signal.

\begin{figure}[htbp]
\begin{minipage}{8.5cm}
\centerline{\psfig{figure=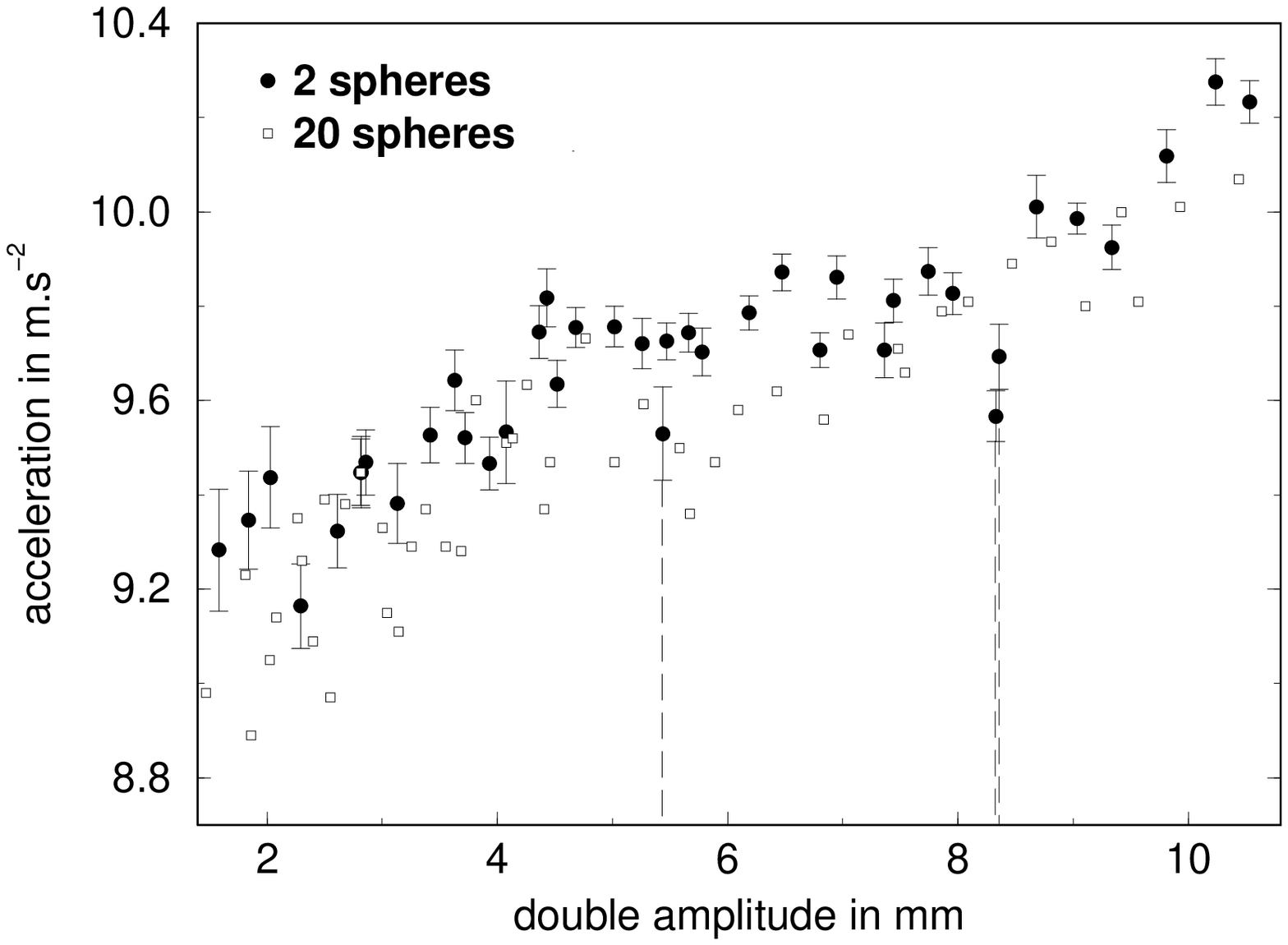,width=8cm,angle=0}}
\centerline{\psfig{figure=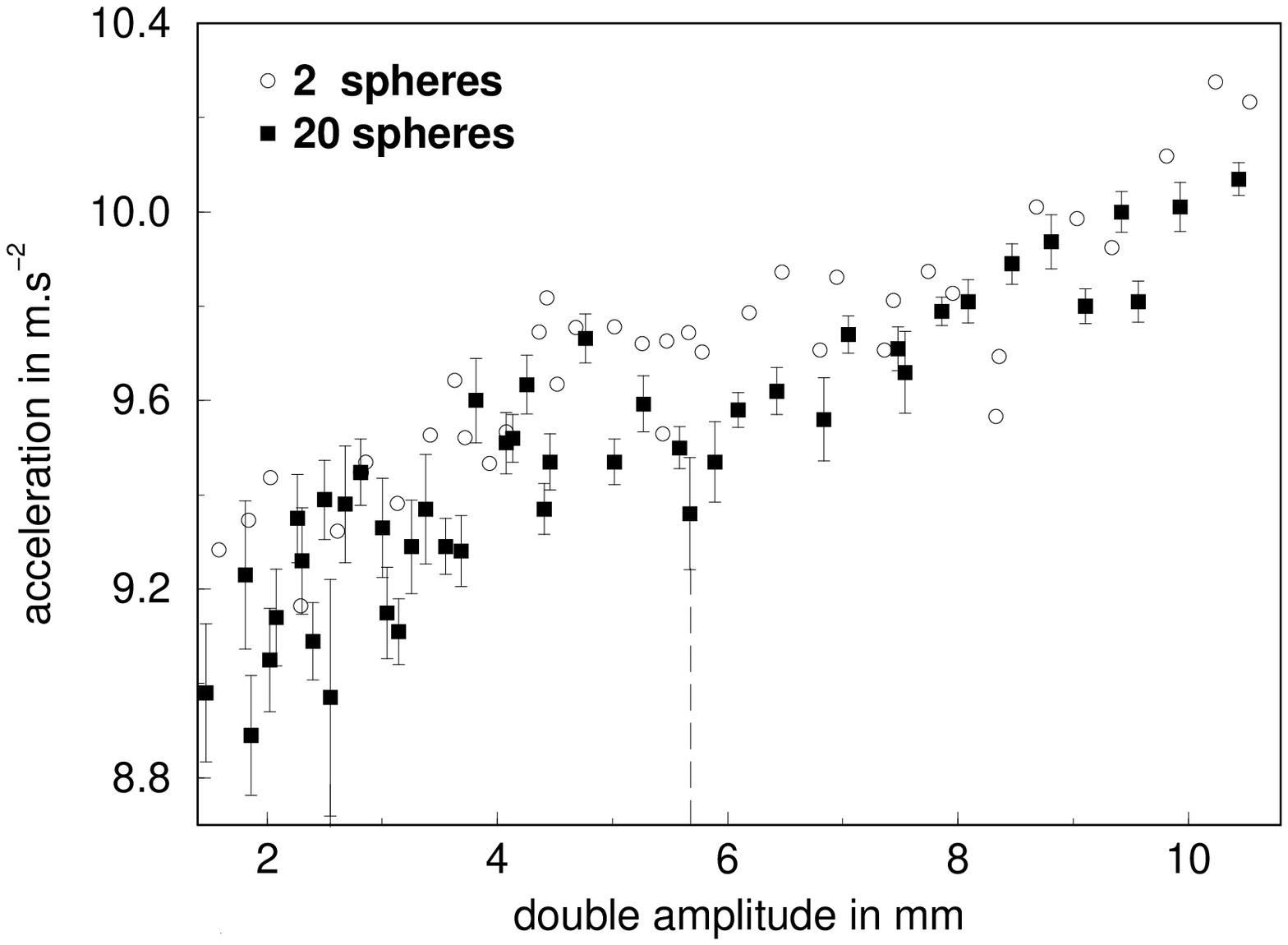,width=8cm,angle=0}}
\caption{All measured data points. In the upper figure error bars 
  have been attached to the measurement for a single free moving bead
  (and another one serving as a well defined lower boundary). In the
  lower part the same data is shown with error bars for the
  measurement using 20 spheres. Data points marked by dashed lines
  have been disregarded for further analysis since they are
  adulterated by resonances of the equipment (see text).}
\label{fig:acc_ampl_2_20}
\end{minipage}
\end{figure}

To collect data points we tuned the adjustable crank\-shaft to a
desired amplitude $A$. Then the motor was accelerated up to a rotation
speed which is slightly less than the speed where we expect
fluidization (initial ramp). At this value all spheres are permanently
in contact (top in Fig.~\ref{fig:cond}). From this point on we
increased the motor speed in very small steps until we observe
separation of the top sphere in each period (bottom in
Fig.~\ref{fig:cond}). This procedure was repeated 10 times for all
amplitudes independently for 2 and 20 spheres. The critical frequency
for each amplitude was determined as the average over the 10
independent measurements. Figure \ref{fig:acc_ampl_2_20} shows the raw
data points with error bars.

The somewhat unusual representation of the data as acceleration vs.
amplitude is due to the fact that our experimental input is the
amplitude, the output, as stated above, is the according critical
frequency and hence the acceleration.

\begin{figure}[htbp]
\begin{minipage}{8.5cm}
  \centerline{\psfig{figure=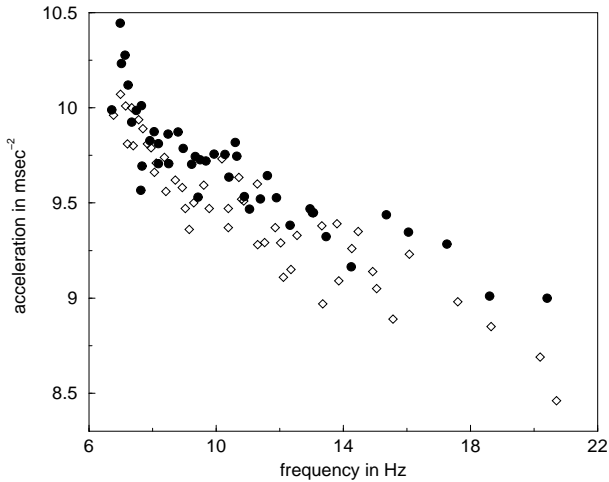,width=8cm}}
  \caption{The critical acceleration versus frequency. Circles 
    represent data of two spheres (reference), diamonds represent data
    for 20 spheres.}
  \label{fig:acc_freq_all}
\end{minipage}
\end{figure}

In Figure \ref{fig:acc_freq_all} the critical acceleration of the
container ($A\omega^2$) vs. frequency $\omega$ for 2 and 20 beads is
shown. One can see that even for two beads the critical acceleration,
which is supposed to be a constant varies significantly with the
frequency, i.e., it decreases with increasing frequency. The curve for
20 beads shows a similar behaviour, however, the critical acceleration
for 20 beads is smaller than the critical acceleration for 2 beads.
This is a first indication to an amplification effect similar as
theoretically predicted.  The experimental device reveals its own
resonances, namely 7.6 Hz for two beads and 11.8 Hz for twenty beads.
These resonances are caused by the mechanical properties of our
experimental setup but not by the properties of the columns of beads
themselves. Data points which belong to these frequencies have been
eliminated for further analysis (see Fig.~\ref{fig:acc_ampl_2_20}).

The data points are affected by large statistical errors, i.e., the
data points scatter around a smooth curve which one expects to find if
only systematic effects were relevant. Therefore, we have to apply a
sophisticated data analysis to extract the amplification effect in
order to compare it with the theoretical prediction. This procedure is
described in the following sections.

\section{Discussion of the Measuring Procedure}

The measurement is affected by different systematic and random errors:
\begin{itemize}
\item Due to the generation of the shaking motion via a connecting rod
  of finite length ($l=18$ cm) there is a systematic deviation of the
  motion from the ideal sinusoidal shaking. Instead of a sinusoidal
  oscillation the generated oscillation is
\begin{equation}
  z=A \cos \phi + \sqrt{l^2-A^2 \sin^2 \phi} -l\,.
\end{equation}
For small ratios $A/l$ we can write
\begin{equation}
  z\approx A\cos\phi - \frac{A^2}{2L}\sin^2\phi
\end{equation}
This leads to a constant shift in the extremal acceleration of the container
\begin{equation}
  \ddot{z}_{Extr}\approx \pm A\omega^2 - \frac{A^2}{L}\omega^2 = 
A\omega^2\left(\pm 1 - \frac{A}{L}\right)
\end{equation}
We see that the absolute value of the acceleration at the upper
turning point is too large by $A/l$ which can be as much as $3\%$.
\item Uncertainty in determining the driving frequency and amplitude
  are very small as described above.
  
\item Due to the noise caused by the motor and the friction of the
  linear bearings the systematic (ideally sinusoidal) vertical motion
  of the column is superposed by a spectrum of high frequencies. These
  waves are source of an undesired extra energy feed which is not
  controllable and causes errors. Measurements of this noise using the
  acceleration sensor have shown that it does not depend much on the height
  of the column, it is determined mainly by the motor and the bearings.
  This property suggests to measure our effect not directly but
  to compare the results for a column of 20 steel spheres with the
  results for a ``column'' of two spheres (see below).

\end{itemize}

\section{Data Analysis}

To eliminate systematic errors we compare the result for a column of
20 spheres with the equivalent measurement for a ``column'' of only
two spheres. The column of two spheres serves as a reference.
Actually, the second bead is fixed to the bottom and oscillates
rigidly while the top bead is free. Effectively there is only one free
moving sphere, for one single sphere taking off from a flat wall could
not serve as a reference for the case when the top bead of a column of
$N$ spheres separates from the rest.  This approach is based on the
assumption that the systematic errors originating from the motor noise
and from the bearings affect both systems almost in the same way.  With the
above mentioned method (see section \ref{sec:setup}) we determined the
critical frequency of driving at which the topmost bead starts to jump
for a given shaking amplitude. From this we calculated the critical
acceleration. Due to the influence of systematic errors also the curve
$a_{\rm crit}(\omega)$ for two beads, which is supposed to be
constant, shows a significant dependence on the driving frequency,
resp. amplitude.  As already mentioned before, in Fig.
\ref{fig:acc_freq_all} one can see that the critical acceleration
decreases with increasing frequency. The curve for 20 spheres shows a
similar behaviour but the critical amplitude is always smaller than
the critical amplitude for 2 spheres at the same driving frequency.
To identify the amplification effect, i.e., to eliminate the error of
the apparatus we divide the critical acceleration for 20 spheres by
the critical amplitude for 2 spheres at the same frequency which
yields the absolute value for the critical Froude number for 20
spheres.  This method is applicable if all significant systematic
errors affect the measurements by a factor which is independent on the
number of beads. This precondition holds certainly for the influence
of the limited length of the connecting rod and holds to a reasonable degree
 for the influence of the higher harmonics in the driving motion due to the
discrete steps in the motor motion. Since we believe these errors to
be the most significant ones our method should give reliable results.

Following this idea we fit both sets of data, for 2 and 20 spheres, to
the function
\begin{equation}
  \label{eq:templf3}
   f(\omega)=\left(a_2\omega^2+a_1\omega+a_0\right)^{-1} 
\end{equation}
and take the ratio of these curves. There are other possibilities to
construct fit-functions which reflect the desired properties of
monotonous decay and significant curvature, it turns out that the
final result does not depend sensitively on the choice of this
function.

\begin{figure}[htbp]
  \begin{minipage}{8.5cm}
    \centerline{\psfig{figure=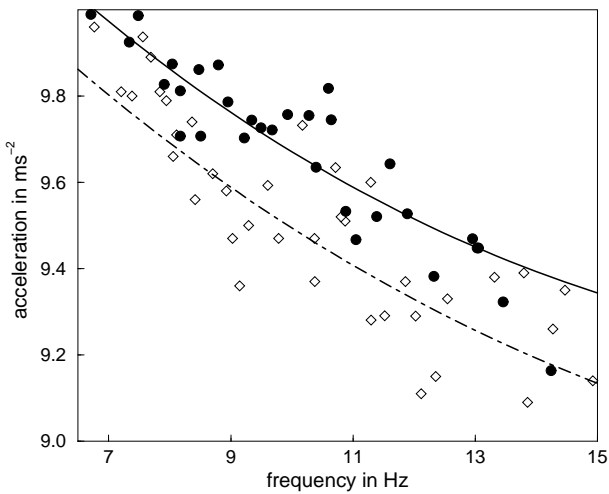,width=8cm,angle=0}}
    \caption{All relevant data points together with a least squares fit 
      curve according to template Eq. (\ref{eq:templf3}). Full
      circles: 2 beads, diamonds: 20 beads. }
    \label{fig:limited_inv_quadr_fit}
  \end{minipage}
\end{figure}

\begin{figure}[htbp]
  \begin{minipage}{8.5cm}
    \centerline{\psfig{figure=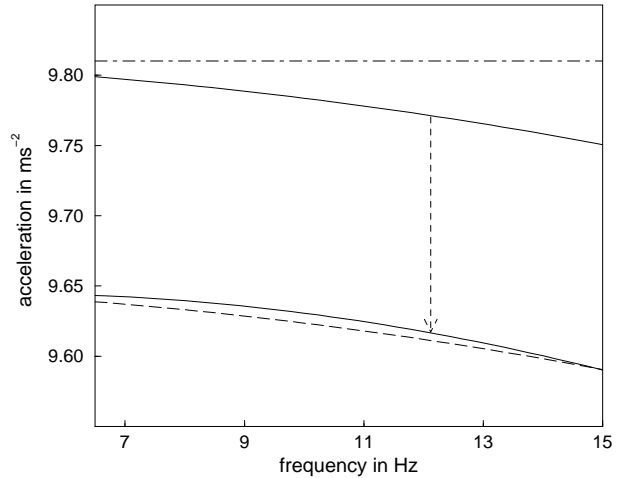,width=8cm,angle=0}}
    \caption{Dashed line: the ratio between the fitted curve for 20 
      and 2 beads (see Fig. \ref{fig:limited_inv_quadr_fit}),
      multiplied by the theoretical value 9.81 m/s$^2$. Solid line:
      theoretical values of the amplification factor given by Eq.
      (\ref{eq.flui.small}) (upper curve) and the same result
      translated vertically for comparison with the experimental one
      (bottom solid curve). The constant shift between both solid
      lines corresponds to about 3\% of the vertical axis due to the
      systematic errors discussed in the text. The dot-dashed line is
      the acceleration of gravity, plotted for reference.}
    \label{fig:final_result}
  \end{minipage}
\end{figure}

\begin{figure}[htbp]
  \begin{minipage}{8.5cm}
    \centerline{\psfig{figure=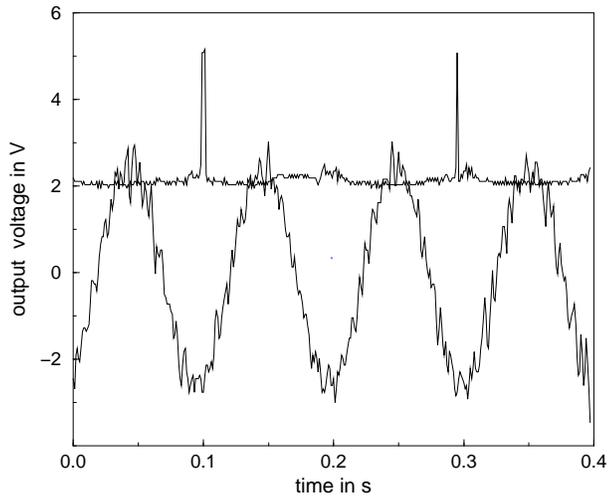,width=8cm}}
    \caption{The output of the acceleration sensor (lower curve). One 
      can see that the sinusoidal curve of the driving frequency is
      disturbed by higher frequency oscillations. The higher
      frequencies account to up to 20\% of the maximum acceleration.
      The upper curve shows the conductivity measurement of the
      topmost beads. The peaks correspond to jumping of the topmost
      bead.}
    \label{fig:oszi}
  \end{minipage}
\end{figure}

The ratio curve, which represents the predicted amplification effect,
can be described in good approximation as a decreasing parabola with
maximum at $\omega=0$. Its curvature is small, in agreement with the
expansion given by Eq. (\ref{eq.flui.small}). Dissipative effects,
represented by the material constant $\alpha$, are not observable at
these frequencies, as the restitution is high for all velocities (0.90
or more): we estimated the second term of $B_4$ being of the order of
10$^{-6}$ or less.  This curve, multiplied by the reference
acceleration 9.81 m/s$^2$, is shown in Fig.~\ref{fig:final_result}.
The ratio between the fitting curves in
Fig.~\ref{fig:limited_inv_quadr_fit} is not exactly 1 at $\omega=0$,
but by about $3\%$ smaller. This discrepancy is due to the above
mentioned systematic errors, namely due to the higher harmonics in the
motion of the stepping motor (see Fig.~\ref{fig:oszi}) which can have
different magnitude at different load. Therefore the division of the
curve for 20 beads by the reference curve alone can not completely
remove this systematic error. Other systematic errors such as
deviations from an ideal sinusoidal motion due to the finite length of
the connecting rod have been removed by the data analysis procedure.

The influence of the higher harmonics can be observed in Fig.
\ref{fig:oszi} showing the output of the acceleration sensor on top of
the column. The sinusoidal motion of the container with driving
frequency $10$ Hz is superimposed by an oscillation of frequency of
about $150$ Hz.  When calculating the maximum acceleration of this
motion one gets a value which is up to 20\% higher than $A\omega^2$.
Thus, by taking $A\omega^2$ as the maximum acceleration of the
container we underestimate the actual acceleration, which, since the
magnitude of the high frequency oscillation varies with different
load, contributes to the difference between the maximum of the ratio
curve in Fig. \ref{fig:final_result} and the theoretical value of
9.81\,m/s$^{2}$.

\section{Discussion}

We investigated experimentally the onset of fluidization of a
vertically vibrated column of spheres. It was found that one observes
fluidization even if the amplitude of acceleration of the vibrating
motion is smaller than the acceleration due to gravity, $A\omega^2/g
<1$. This result is in agreement with the theoretical prediction
\cite{BelowG}. The quantitative theoretical result for the critical
parameters for the onset of fluidization given by Eqs. (\ref{eq:flui})
and (\ref{eq.flui.small}) does not contradict the experiment but could
also not be conclusively confirmed by the experiment due to
insufficiencies of the experimental setup. The most important
shortcoming is the existence of high frequency vibrations mainly due
to the motor noise and the bearings. While qualitatively agreement is
shown, due to these limitations the experimental measurements are not
completely conclusive to verify the theoretical results: a
quantitative check of the theoretical predictions requires a more
sophisticated experimental setup, directed to reduce noise and enhance
the effect. The noise can be reduced by improving mechanical
isolation: enclosing the column of spheres into a massive block or
wall, using very rigid bearings and separating the motor at a good
distance from the column. Higher columns or more balls will help to
reduce the noise, but one has to work harder on a correct vertical
alignment; for this reason the beads should not be very small. The use
of a softer material alone would enhance the effect, but then one
would like still good restitution, smooth surface and good
conductivity (for the electrical method of determination), so the
choice of material is not straightforward. Also higher frequency will
enhance fluidization at lower Froude numbers, but then controlling
amplitude (and noise) will become much more critical.

\medskip 

\begin{acknowledgement}
  We thank Hans Scholz for discussion and Christine Rosinska for
  technical aid.
\end{acknowledgement}

\end{document}